\documentclass[ reprint, amsmath,amssymb, aps,pre,superscriptaddress]{revtex4-2}

\usepackage{graphicx}
\usepackage{dcolumn}
\usepackage{bm}

\begin{document}

\title{The Dynamics in Vibro-fluidized Beds: A Diffusing Wave Spectroscopy Study}

\author{Marlo Kunzner}\affiliation{Institut für Materialphysik im Weltraum, Deutsches Zentrum für Luft- und Raumfahrt (DLR), 51170 Köln, Germany}

\author{Christopher Mayo}\affiliation{Institut für Materialphysik im Weltraum, Deutsches Zentrum für Luft- und Raumfahrt (DLR), 51170 Köln, Germany}

\author{Matthias Sperl}\email{Matthias.Sperl@dlr.de}\affiliation{Institut für Materialphysik im Weltraum, Deutsches Zentrum für Luft- und Raumfahrt (DLR), 51170 Köln, Germany}\affiliation{Department for Theoretical Physics, University of Cologne, Germany}

\author{Jan Philipp Gabriel}\email{Jan.Gabriel@dlr.de}\affiliation{Institut für Materialphysik im Weltraum, Deutsches Zentrum für Luft- und Raumfahrt (DLR), 51170 Köln, Germany}

\date{\today} 

\begin{abstract}
We demonstrate the densification of a granular model system of polystyrene spheres over time by shaking with varying excitation amplitudes or effective temperatures. This densification is quantified by the mean square displacement (MSD), which is measured by diffuse wave spectroscopy (DWS) of a sinusoidally excited vibrating fluidized granular bed. The DWS method also extracts the inherent heterogeneous dynamics of the system in the bulk and at the wall. Through an empirical model-based extraction we obtain the ballistic and diffusive time constants, as well as caging sizes, which were found to depend on temperature and density. The results obtained from this study reveal a sub-diffusive power-law behavior in the MSD, indicating an arrest of motion and potentially a glassy system, especially in cases where the excitation is low to moderate compared to gravity. The extracted MSD caging sizes are two orders smaller than the Lindeman length found in colloidal systems.
\end{abstract}

\maketitle

\section{\label{sec:Intro}Introduction}
In everyday life, it is common to observe granular matter in motion. The movement of wind, for example, can be seen in the drifting of sand dunes or the occurrence of sandstorms. Within industrial contexts, another example of granular matter in motion is the passage of wheat grains through a silo. The perspective of material science addresses the diverse forms of motion within granular systems, which bear resemblance to the solid, liquid, and gas phases characterized in thermodynamics. \cite{andreotti2013granular,duran2012sands,rosato2020segregation}. A primary area of interest is the transition from liquid-like states to solid-like states, which can be described as a glass transition \cite{liu1998jamming,berthier2009glasses}. This transition, as well as the dynamics of granular media in general, are in some respects more complicated than in atomic matter, due to the many interactions such as electrostatics, friction, dissipation, and gravity \cite{andreotti2013granular,duran2012sands,rosato2020segregation,barabasi1999physics,tomasetta2014correlation}. Many macroscopic descriptions of the behavior of powders exist, but looking at inter-particle dynamics is often complicated, as powders are dense and opaque \cite{andreotti2013granular,rosato2020segregation,ehrichs1995granular,wildman2001granular}. Here we consider a granular model system of polystyrene spheres driven by sinusoidal oscillating agitation. To provide a good statistical description of the system and take advantage of the opaqueness, we use Diffusive Wave Spectroscopy (DWS) as a method to measure intensity correlation functions and then subsequently calculate the mean-square displacements (MSD) of our particles under certain assumptions \cite{berne2000dynamic,furst2017microrheology}. DWS has previously been used to study gas fluidized beds and granular flows \cite{menon1997diffusing,menon1997particle,M12}. However, to the best of our knowledge, at the time of this publication, no study has been published investigating vibro-fluidized beds with DWS. The present paper explores the influence of vibro-fluidized beds on the dynamics of granular media. It does so by utilizing the key features of DWS, mainly the ability to measure high spatial and temporal resolution with regard to a large number of particle movements. This in turn enables a comparison of this system to funnel and gas-fluidized beds. The findings from this study will serve as a solid foundation for future research, including experiments conducted under microgravity conditions on the International Space Station, as well as extending the investigation to different agitation methods, particle sizes, and frictional properties.

\section{\label{sec:ExpPro}Methods}
\noindent
Shaken granular matter is commonly characterized by comparing the gravitational acceleration with the vibrational acceleration using the dimensionless acceleration \cite{rosato2020segregation,evesque1990surface,douady1989subharmonic,eshuis2007phase}
\begin{equation}
    \label{eq:DimlessAcc}
    \Gamma = \frac{A\,\omega^2}{g}.
\end{equation} 
with $A$ the oscillation amplitude, $\omega$ the angular frequency, $g$ the gravitational acceleration, and $\Gamma$ the dimensionless acceleration. For $\Gamma < 1$ particles should not be excited enough to be lifted up or slide along each other, but rather show small oscillations. However, it is observed that real systems with shear show densification \cite{rosato2020segregation,Fauve1989convective,eshuis2007phase}. Whereas for a critical $\Gamma_c$ above 1 ($\Gamma > 1$), the particles show relative motion and can no longer densify \cite{rosato2020segregation}. 
Shaking a granular packing of frictional spheres will inevitably lead to shear, making it important to note the existence of dilatancy. Shearing a granular packing above a density threshold will make it expand, if below the threshold the packing will compact \cite{schroter2017local,andreotti2013granular}. 
\begin{figure}[t!]
    \centering
    \includegraphics[width=0.8\linewidth]{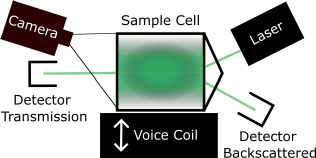}
    \caption{Schematic of the set-up with a sample cell mounted on a voice coil for excitation and a laser source. Light is collected in transmission and backscattering geometry and then fed into Avalanche Photo Diodes (APDs) and a correlator.}
    \label{fig:Setup}
\end{figure}
\noindent
For tracking the particle movements we use DWS as a non-intrusive scattering technique \cite{BrownDynamic,weitz1993diffusing} illustrated in Fig. \ref{fig:Setup}. The incident laser light undergoes multiple scattering events in the sample before being detected. The fluctuations contain the averaged dynamics of the observed scattering volume and can be correlated to obtain the Intensity Correlation function $g_2(t)$, defined as
\begin{equation}
g_{2}(t) = \frac{\langle I(0) I(t) \rangle}{\langle I(0) \rangle \langle I(t) \rangle}.
\label{eq:oszimodel}
\end{equation}
with intensity $I$. The voice coil used to excite the system and the sinusoidal excitation are shown in Fig. \ref{fig:Setup}. The oscillation $O(t)$ is modifying the intensity correlation function  $g_2(t)$ as follows
\begin{equation}
g_{2,\textbf{mes}}(t)-1 = O(t) (g_2(t)-1) 
\label{eq:oszimodel2}
\end{equation}
with
\begin{equation}
O(t) = \frac{1}{T} \int_0^T \exp( -\kappa^2 A^2 \sin(\omega(t + t')) - \sin(\omega t') )^2 dt'
\label{eq:oszi}
\end{equation}
and $T$ as period, and $\kappa$ the variance of the scattering vector (inspired by \cite{M12}). The dynamics of the particle are related to the fluctuating electric fields. Assuming the Siegert-Relation 
\cite{berne2000dynamic}.
\begin{equation}
    \label{eq:Siegert}
    g_2(t) = 1 + \Lambda\, |g_1(t)|^2
\end{equation}
the intensity correlation function $g_2(t)$ is related to the electric field correlation function $g_1(t)$ (FCF)\cite{berne2000dynamic}. With the coherence area factor $\Lambda$. 
Assuming the DWS approximation for the electric field correlation function $g_1(t)$ that the light is undergoing a random walk \cite{furst2017microrheology} we can relate it to the mean square displacement (MSD) $ \langle \Delta r^2 \rangle$ by
\begin{equation}
    \label{eq:FieldAutoCorr}
    g_1(t) = \exp(-\frac{1}{3}\left(\frac{kL}{l^*}\right)^{2} \langle \Delta r^2 \rangle)
\end{equation}
with $k$ the wave vector of the incident light with a magnitude of $1.18\cdot10^9$ 1/m, $L$ the effective length of the sample, in transmission geometry, the sample length of $5$ mm, and $1.43$ mm in backscattering geometry \cite{Brown1993Dynamic}. The randomization length $l^*$ is calculated (see eq. \ref{eq:lstar} to \ref{eq:Rho}) as $480$ µm, and can be viewed as the distance until the direction of light propagation is randomized, $r$ is the 3-D distance a particle moves, and $\langle\,\rangle$ indicates an ensemble average \cite{furst2017microrheology,berne2000dynamic}. The randomization length is given by 
\begin{equation}
    \label{eq:lstar}
    l^* = \frac{l}{1-\cos(\theta)}
\end{equation}
with the scatterer properties as the distance between scattering events $l$ and scattering angle $\theta$. The experimental determination is difficult \cite{utermann2012friction}. The distance between scattering events itself is proportional to the inverse of the number density of the scatterers, or volume fraction, ($\rho_N$) times scattering cross-section 
\begin{equation}
    \label{eq:lzero}
    l(\sigma) = \frac{1}{\rho_N\,\sigma}
\end{equation}
depending on the ratio of the wavelength of the incident light to the particle size, according to Mie-Theory \cite{mie1908beitrage} and calculated by python program PyMieScatt \cite{sumlin2018retrieving}. In our case the particles' radii are larger than the wavelength of the light, but not sufficiently large to consider Fraunhofer or geometrical optics. According to Mie-Theory, backscattering and forward scattering will have different intensities. The volume fraction is given by
\begin{equation}
    \label{eq:Rho}
    \Phi = \rho_N = \frac{4\,\pi}{3}\cdot \frac{a^3\,N}{V_{Cell}} 
\end{equation}
with particle number $N$ and spherical particles with radius $a$ occupying a cell volume $V_{Cell}$.
\begin{figure}[th!]
    \centering
    \includegraphics[width=1\linewidth]{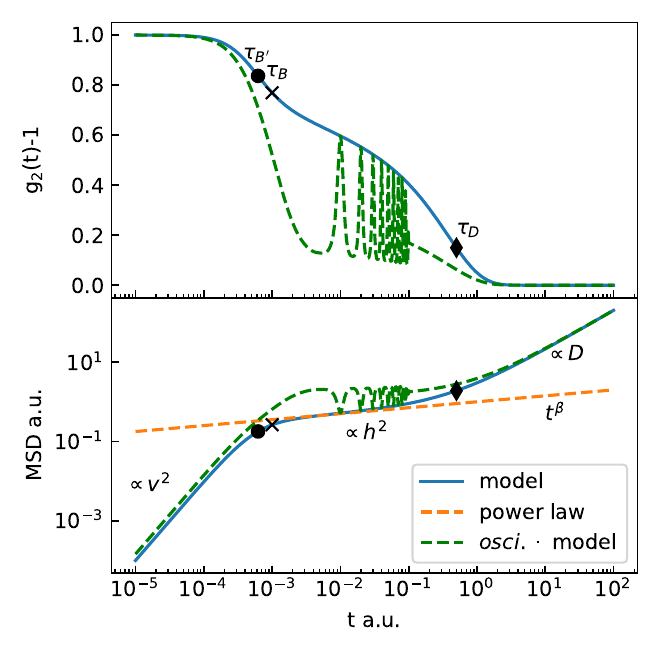}
    \caption{Depiction of Eq. \ref{eq:JanFit} at arbitrary lag times with a superimposed shaking oscillation. The upper graph shows $g_2(t)$ versus time plot and the lower one shows the MSD. The dashed green line is a measurement including an oscillations and the blue line is the model function. The orange line depicts a power law without diffusion.}
    \label{fig:MSDFunc}
\end{figure}
The particles' dynamics can now be extracted by empirically modeling the dynamics with power laws in the MSD representation for the three expected behaviours ballistic, caging, and diffusion by
\begin{equation}
    \label{eq:JanFit}
    \langle \Delta x^2 \rangle = \frac{1}{ (\tau_B/t)^2+ h_D^{^-2}t^{-\beta}}   +   \frac{t}{\tau_D}+\frac{\tau_B}{\tau_D} \left(\exp\left(-\frac{t}{\tau_B}\right)-1\right)
\end{equation}
with the diffusion timescale $\tau_D$, ballistic timescale $\tau_B$, and height $h_D$ related to the caging power law, taken where the diffusion behavior of the system appears. The fist term in equation \ref{eq:JanFit} is a ballistic and caging power law, the second term is the diffusion power law and the third term is correcting term for the non additive contribution of the ballistic and caging power law with the diffusion power law. The measured intensity correlation function $g_{2,\textbf{mess}}$ can be calculated by applying the model Eq. \ref{eq:JanFit} on Eq. \ref{eq:FieldAutoCorr} and using Eq. \ref{eq:oszimodel} to obtain the result shown as the green dashed line in Fig. \ref{fig:MSDFunc}. The blue line describes the particle dynamics by $g_2(t)$ without the oscillations. We use this function to extract the velocity at short time scales, the supposed caging length at intermediate time scales, and diffusing behavior at long time scales. For an effective ballistic time $\tau_{B'}$ 
\begin{equation}
    \label{eq:tauB}
    \tau_{B'} =  t(\max( \langle \Delta r(t)^2 \rangle / t))
\end{equation}
and effective ballistic length scale $h_{B'}$
\begin{equation}
    \label{eq:hB}
    h_{B'} = \sqrt{ \langle \Delta r(\tau_{B'})^2 \rangle}
\end{equation}
Following that DWS averages over relative motions of the entire sample volume since the interference pattern is made of components from everywhere in the sample, the MSD values are averaged and measured velocities are the mean velocities of fluctuating particles. 
The velocity fluctuation from the MSD 
\begin{equation}
    \label{eq:MSDVel}
    \lim \limits_{t \to 0} \langle \Delta r^2 \rangle = \langle \delta v^2 \rangle t^2 = \overline{v}^2 t^2.
\end{equation}
is related to the granular temperature \cite{biggs2008granular}
\begin{equation}
    \label{eq:GranTemp}
    T_{Gran_i} = m \cdot\overline{v_i^2} = \frac{1}{3}\,m\cdot \overline{v}^2
\end{equation}
with $m$ the particle mass and $v_i$ the velocity component of one of the three directions in 3-D space \cite{menon1997particle,wildman2001granular,andreotti2013granular}.

Furthermore, with the velocity we can estimate the possible deformation and collision time of the particle. For two spherical particles forming and breaking contact, the surface deformation and collision times can be calculated according to Hertz \cite{Hertz1881Contact,brilliantov1996model,johnson1982one}. The first important parameter is the elastic modulus ($E$) of the two spheres colliding, giving the parameter $\Lambda \propto 1/E$ \cite{brilliantov1996model}. Using $\Lambda$ and the radius of the particle $a$ the spring constant of the contact can be calculated in Eq.\ref{eq:HertzSpringConst} \cite{brilliantov1996model}. 
\begin{equation}
    \label{eq:HertzSpringConst}
    k^2 = \bigg(\frac{4}{5}\frac{2}{3\Lambda}\bigg)^2 a
\end{equation}
Following this we can use the velocity calculated by DWS ($\overline{v})$, the mass of a particle ($m_p$), and Eq.\ref{eq:HertzTimeFull} to calculate the contact time \cite{brilliantov1996model}.
\begin{equation}
    \label{eq:HertzTimeFull}
    t_c = 2.94 \left(\frac{m}{\sqrt{\left(\frac{4}{5}\frac{2}{3\Lambda}\right)^2\cdot R }}\right)^{2/5}\overline{v}^{-1/5} = 2.94 \left(\frac{m}{k}\right)^{2/5}\overline{v}^{-1/5}
\end{equation}
Afterward, we can place the indentations caused by the collision in the MSD graphs according to $\delta^2 = (t_c\cdot\overline{v})^2$.

\subsection*{\label{sec:Expdetails}Experimental details}
The set-up comprises a 15x15x5 mm$^3$ sample cell containing 0.602 $g$ of 140 $\mu$m polystyrene particles (Dynoseeds 140) mounted on a voice coil (Visaton EX 80 S) and a 532 nm laser (Coherent Verdi G5 SLM), as is shown in Fig. \ref{fig:Setup}. The container is shaken via an oscillation from the voice coil. This supplies the system with energy, thereby agitating the system. The green laser light is scattered multiple times and detected via single-mode fibers and polarisers connected to avalanche photodiodes (APDs) in backscattered and transmission geometry. The signal is then passed into a hardware correlator (ALV 7004). The sample, being mounted on a voice coil, can move vertically via the voice coil's sinusoidal oscillation. In addition, a camera (Panasonic hc-v180) is mounted to perform video surveillance to collect information on the packing fraction. The image analysis is done using Fiji (ImageJ) \cite{schindelin2012fiji}.\\
The experimental procedure is as follows:\\
We vacuum-dry the sample, weigh it, and fill it into the sample cell. Afterward, we start the camera recording, and following that the shaking is started with a constant frequency of 100 Hz at a chosen $\Gamma$ value by tuning the amplitude, almost instantly after the shaking is started the DWS collects data. For reproducibility, we start with a strong excitation to always create a loose system. The system is analysed for 10 seconds with the measurement time doubling iteratively until 5120 seconds.\\
The time steps are chosen for two reasons: Firstly: DWS needs many scattering events for proper statistics. The statistics become insufficient in the highest 3 decades measured, thereby measuring at least 1000 times as long as the highest timescale we want to observe is necessary \cite{berne2000dynamic}. Secondly, the system changes over time and the time steps give insight into the dynamics at different volume fractions of the system.

\section{\label{sec:DatAna}Results}
\subsection{\label{sec:PackFrac}Packing Fraction}
\begin{figure}[t!]
    \centering
    \includegraphics[width=\linewidth,keepaspectratio]{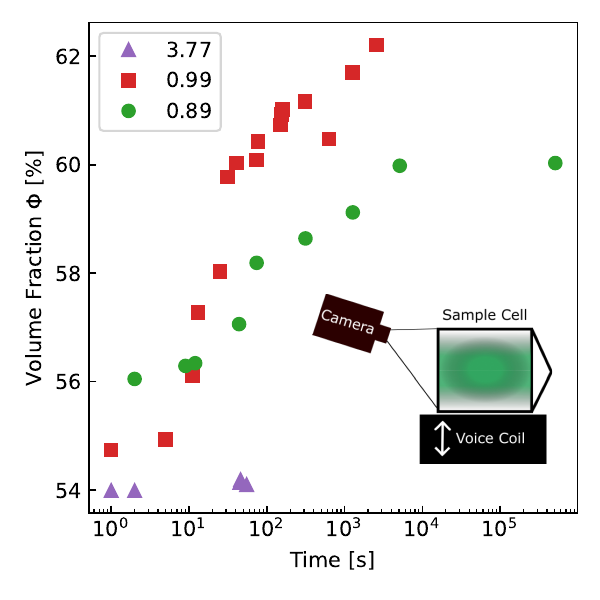}
    \caption{Evolution of the volume fraction for different $\Gamma$ values. The volume fraction in \% is plotted versus time to observe any densification. Shaking up is shown in purple triangles.}
    \label{fig:AllPFEvo}
\end{figure}

The packing fraction is determined by using video microscopy during shaking. Determination of the packing fraction using this method can show densification throughout the experiment, which is further supported by changes in the correlation functions seen in Sec. \ref{sec:Dyn}. Fig. \ref{fig:AllPFEvo} shows the relationship between packing or volume fraction and time. As expected the system densifies over time for both $\Gamma$ values around 1. The shaking up ($\Gamma = 3.77$) 
does not further compact the sample. The plateau for $\Gamma = 0.89$ can be explained by the dilatancy onset for mono-disperse spheres at $\Phi_D\,\approx\,60\,\%$ depending on friction and pressure and this $\Gamma$ value does not supply enough energy to rearrange the beads and further densify the system \cite{schroter2017local}. In contrast, the system becomes denser with $\Gamma = 0.99$ as agitation continues. We base this on the same reasoning, as the particles should be able to move relative to each other at $\Gamma_c$ and become somewhat fluid-like. This can densify the system even above the threshold of $\Phi_D\,\approx\,60\,\%$.  

\subsection{\label{sec:Dyn}Dynamics}
\begin{figure}[t!]
    \centering
    \includegraphics[width=0.95\linewidth,keepaspectratio]{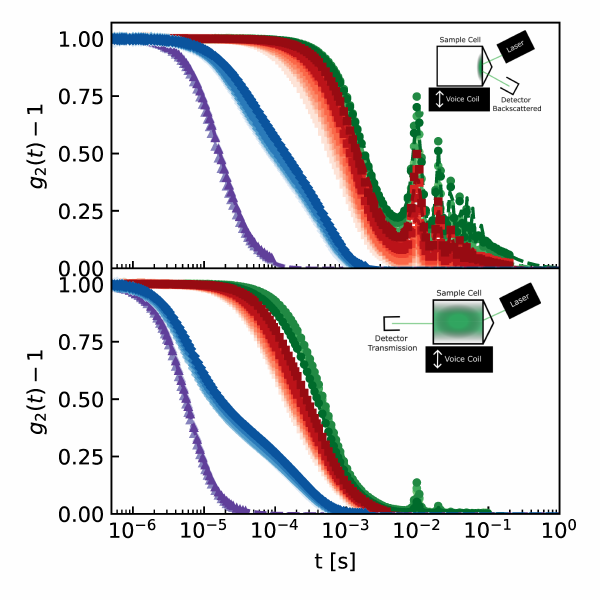}
    \caption{$\Gamma$ and time-dependent Correlation functions $g_2(t)$ in backscattering (top) and transmission (bottom) geometry, with figure inserts depicting the probed volume. The $\Gamma$ values of 3.77, 1.1, 0.99, and 0.89 are shown by the purple, blue, red, and green curves respectively. The color scheme is designed such that darker shades of the color indicate later or more dense versions of the same system.}
    \label{fig:corr}
\end{figure}

Fig. \ref{fig:corr} shows $\Gamma$ and time-dependent correlation functions $g_2(t)$ in backscattering (top) and transmission (bottom) geometry. The purple, blue, red, and green curves show $\Gamma$ values of 3.77, 1.1, 0.99, and 0.89 respectively. Recorded data is shown as symbols and the fit as lines, showing a good agreement between the fit function and data. The set-up depictions in the graphs indicate the probed scattering volume. The oscillations look more pronounced in the backscattering correlation functions and the time is shorter for transmission geometry across all measurements to fully decay. Both factors are not necessarily representative of stronger underlying oscillation amplitudes as well as faster or slower underlying dynamics respectively, which will be clearer after the discussion in \ref{sec:Analysis}.

\section{\label{sec:Analysis}Analysis}
\begin{figure}[t!]
    \centering
    \includegraphics[width=0.95\linewidth,keepaspectratio]{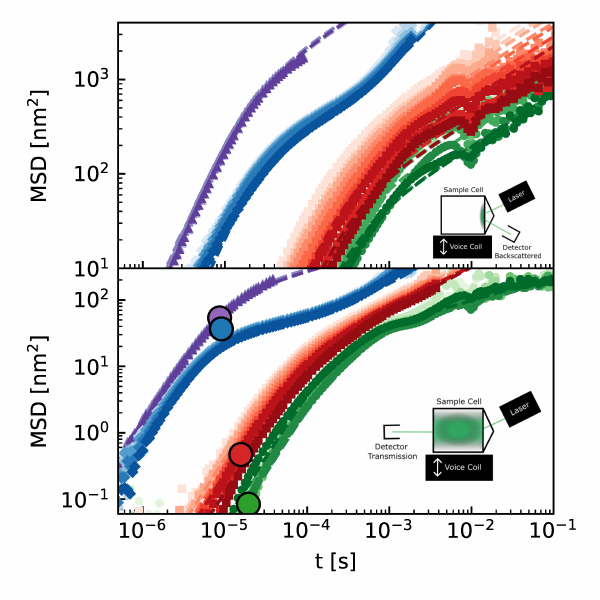}
    \caption{$\Gamma$ and time-dependent MSDs calculated from data in fig \ref{fig:corr} in backscattering (top) and transmission (bottom) geometry and model fits following equation \ref{eq:JanFit}. The $\Gamma$ values of 3.77, 1.1, 0.99, and 0.89 are shown by the purple, blue, red and green curve respectively. The color scheme is the same as previously. The dots in the lower graphs show the lengths and times calculated by assuming Hertzian contacts.}
    \label{fig:MSD}
\end{figure}

Fig. \ref{fig:MSD} shows $\Gamma$ and time-depended MSDs calculated from data in Fig. \ref{fig:corr} shown as symbols and model fits following equation \ref{eq:JanFit} shown as dashed lines. Since the model can describe the agitation oscillation, we were able to remove the oscillations present in the MSD representation to an almost oscillation-less depiction. As illustrated in Fig. \ref{fig:MSD}, the oscillations observed in the correlation functions, which exhibit significant differences, demonstrate notable similarity when translated into the MSD plots. The measured MSDs reveal the presence of power law behavior, initially manifesting as a power law exponent of 2 at very short time scales, indicative of ballistic motion or elastic deformation. Subsequently, a transition occurs into a plateau-like interval, which may represent caging behavior. The dots in the lower graph indicate collision times and indentations calculated by the Hertzian model, while the blue curves ($\Gamma$ = 1.1) additionally show a power law exponent of 1, which is indicative of diffusive motion. For progressively more dense systems, like the red curves, the diffusion is at a later time and has higher displacement values than the technique can resolve. Conversely, the purple curves show no diffusion since the strong agitation has ballistically displaced the particles too far to resolve any particle interactions seen by a full decorrelation before any plateau has formed.
\begin{figure}[t!]
    \centering
    \includegraphics[width=0.95\linewidth,keepaspectratio]{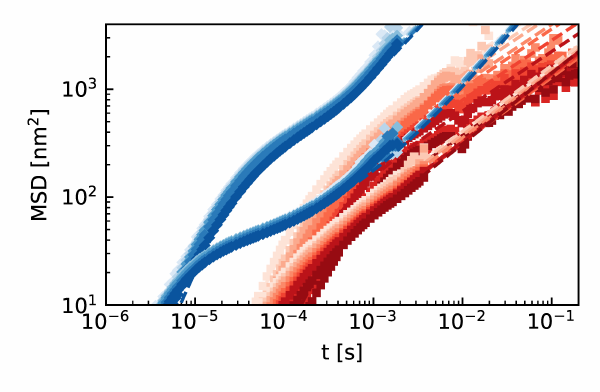}
    \caption{$\Gamma$ and time-dependent MSDs in both geometries joined for $\Gamma$ values of 1.1 and 0.99, where the higher MSDs values indicate backscattering geometry.}
    \label{fig:Fig6}
\end{figure}
To highlight the differences in traveled distances Fig.\ref{fig:Fig6} shows a joined graph of 1.1 and 0.99 $\Gamma$ in backscattering and transmission geometry. It is demonstrated that the backscattering and transmission geometry exhibit a high degree of similarity in shape, with a divergence occurring at the point where the ballistic regime ends. We attribute this to inhomogeneities in the sample. The transmission curves have a lower MSD value than the backscattering curves, suggesting a shorter particle displacement traveled. We expected that the transmitted light experience a greater exposure to the bulk of the material, thereby providing an average dynamics of the entire sample. Meanwhile, the backscattered light probes the sample close to the walls. This point is highlighted by the small sample depictions in the figures, where the green color represents the scattered light.
\begin{figure}[t!]
    \centering
    \includegraphics[width=0.95\linewidth,keepaspectratio]{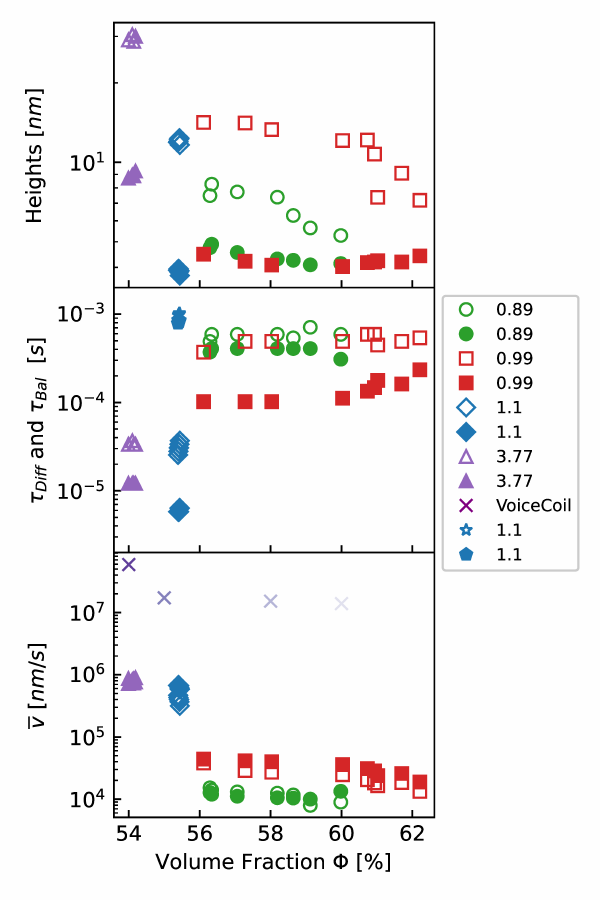}
    \caption{Extracted correlation times for ballistic motion $\tau_B$ and diffusive motion $\tau_D$ as well as ballistic particle velocity $v$ and plateau values $h$ by fitting of equation \ref{eq:JanFit} to data in Fig. \ref{fig:corr}. Empty symbols for backscattering geometry and full symbols for transmission geometry. The star and pentagon show the diffusion timescale. }
    \label{fig:pararho}
\end{figure} 
The packing differs close to the wall compared to the bulk since the wall prevents proper close packing.\\
The extracted fit parameters are plotted against volume fraction in Fig. \ref{fig:pararho}, with empty symbols for backscattering geometry and full symbols for transmission geometry. The fit parameters $h$ and $\tau_{Bal}$ are lower for transmission than for backscattering geometry. The plateau heights increase with higher excitation energies, as shown Fig. \ref{fig:pararho} a). As the system's density increases, the particles enter the caging regime earlier, a phenomenon that can be quantified by observing the decrease in height. Ballistic time scales are visible in Fig. \ref{fig:pararho} b) and show a slight increase as the system densifies for $\Gamma < \Gamma_C$, indicating that the system is becoming slower. The diffusive time scales $\tau_{Diff}$ only visible in the blue MSD curves, are shown by the star and pentagon symbols. The velocities, shown in Fig. \ref{fig:pararho} c), are calculated according to Eq. \ref{eq:MSDVel} and shows the expected behavior, as it increases with excitation energy and additionally decreases with density which is shown by the red and green symbols.

\section{\label{sec:Discusion}Discussion}
The system densifies over agitation time, as shown in Fig. \ref{fig:AllPFEvo}, this is to be expected as vibrations are a commonly used and highly effective way to densify granular media \cite{knight1995density,nowak1998density}. It can also be seen that the system stops densifying at the dilatancy on-set for excitations below $\Gamma = 1$. This phenomenon can be attributed to the inherent difficulty of particle movement relative to each other, thereby hindering the potential for reorientation-driven densification. Higher excitations do not show this property, especially since $\Gamma = 3.77$ visibly fluidizes the sample and has a measured volume fraction below 54\%, which is close to the most loose solid  granular packing \cite{schroter2017local,farrell2010loose}.\\
We also observe a clear difference between backscattering and transmission measurements for excitations close to and above $\Gamma = 1$. This leads us to conclude that the system is strongly heterogeneous. At low $\Gamma$ values this behavior is less prominent. We suspect convection rolls to be one of the causes of this effect, confirmed visually via the video taken of the system \cite{wildman2001granular}. Due to the nature of the scattering geometry, the transmission measurements provide us with a more complete insight into the average behavior of the bulk, as the scattering path passes through the whole sample. This has the benefit of reducing the effect of localized velocity differences in the sample.\\
From the fits of the MSD curves in Fig. \ref{fig:MSD}, we extract parameters that align with what the MSD curves show by visual inspection. The power law exponent $\beta$, indicative of the slope of the MSD after the ballistic regime, has a constant value of 0.55 for backscattering geometry but differs between excitations for transmission. A $\beta$ of 0.55, called sub-diffusive motion, is not uncommon for colloids in solution \cite{Hunter2012Colloid}. This leads to the assumption that the observed behavior is analogous to a thermal system and thus may indicate caging behavior. \cite{van2017cage,Hunter2012Colloid,xing2018microrheology,cardinaux2002microrheology,Fuchs2016Glass}. In general an exponent of 0.5 is found in strongly collective dynamics for molecular substances and is assumed to be a generic feature of structural relaxation \cite{pabst2021generic}. The power law has slight variations between measurements this could be caused by the small amount of polydispersity of our system as colloidal DWS measurements with mono and polydisperse particles show similar characteristics \cite{furst2017microrheology,van2004brownian,menon1997particle}.
Estimation shows that this behavior is influenced by the frictional forces between the particles, and that tuning the friction would change the power law exponent \cite{schroter2017local,Bowden1982Friction,gao2009experimental}.\\
The MSD is cut at the resolution limits given by $k_0$, mean distances larger than $k_0$ can not be resolved. The calculated $T_{Gran}$ and the limits align well with literature over several decades \cite{menon1997particle,menon1997diffusing}. The resolution limit, however, is not only given in distance but also in time. This leads us to two likely explanations of the cause of the initial ballistic decorrelation. DWS is a statistical average of all dynamics in the scattering volume. As our system has several thousand particles probably in movement, each one of them needs to move only a fraction of what is given for the one-event decorrelation case.\\
Firstly, we consider the clapping contacts mentioned in literature on granular media \cite{tournat2004probing,van2013evolution}. Here contacts would form due to collisions these would then deform the particles up 10 nm, when assuming Hertzian contacts, and break apart again \cite{rosato2020segregation,Hertz1881Contact,johnson1982one,mathey2024device,brilliantov1996model}. The collision process takes roughly $5\cdot 10^{-6}$s for our particles at the highest agitation, slight deviations are depending on the excitation energy \cite{brilliantov1996model}. It should be noted that the time scale relating to each contact time for each agitation aligns with the ballistic-like increase in the MSD and continues even after the Hertzin contacted is expanded. The behavior appears ballistic since the resolved Hertzian part of the MSD is linear following Hooks law, and non linear deviations from ballistic behavior should be on even shorter unresolved times. Additionally, the indentations created by making that contact align with the MSD values, making it plausible that the short time scale behavior analyzed is the contact formation breaking of the Hertzian contact. Lower excitations lead to a bigger deviation from the Hertzian model since the particles are no longer breaking apart, but follow the movement of the voice coil. Interestingly data of Scalliet et al. \cite{scalliet2015cages} for very high excitations can be used to extrapolate our data, showing that the actual ballistic behavior happens at slightly larger time and length scales than probable by DWS with green light \cite{scalliet2015cages}.\\
Secondly, it could be either translational or rotational motion of the particles or a combination of both. As the movement on a nanometer scale is very small compared to the size of our particles, one could also attribute the decorrelation to rattling or vibrations \cite{rosato2020segregation}. 1 Å is also equal to 0.01\% of a particle's circumference implying that the amount one particle would need to turn to decorrelate the light path is achievable, even at short timescales.\\
The surface is assumed to be a strong scattering source, giving us information about the motion of the particle surface (leading to sensitivity for deformation, translation, and rotation). Since the largest refractive index gradient is at the air-particle boundary, the light is scattered weaker on the particle length scale relative to the surface roughness. This can be further illustrated by the observed brightness of each sphere compared to other materials\cite{rysselberghe2002remarks,Nayar1989surface} and lead to a slide underestimation of l$^*$ resulting in smaller MSDs than assumed.\\
Measured particle velocities are lower than the velocity of the voice coil, as they should be when considering dissipation, consequently, the excitation itself is not responsible for decorating the light. The velocities for high voice coil excitations agree with measurements for granulates falling through a funnel, as both show a mean velocity fluctuation of 0.1-1 $mm/s$. \cite{menon1997diffusing,menon1997particle}. We measure MSD plateau values between $3\cdot10^1$ and $3\cdot10^3 nm^2$ for transmission and backscattering respectively. The funnel experiment shows a plateau value of roughly $10^3 nm^2$ for particle sizes of $95 \mu m$ and $194 \mu m$, covering larger and smaller particle sizes than ours \cite{menon1997diffusing}. In the gas-fluidized bed, the plateau is a little below $10^4 nm^2$, this intuitively makes sense as a gas-fluidized bed will be less dense than our system and the particles should have a higher mean free path \cite{menon1997particle}. We like to emphasize that the funnel experiment resolves the beginning of diffusion consistently with applied camera-extracted diffusion values underlining the reliability of the extracted length-scales \cite{menon1997particle}.
Our particles show no sign of bubbling, therefore the system would be classified as Group A, according to Geldart \cite{geldart1973types}. Menon's particles, used in a gas-fluidized bed, would either be class A or B \cite{menon1997particle}. This leads us to believe that the Geldart classification for fluidized beds should be able to be transformed into a granular temperature-dependent version.\\
When comparing Mode-Coupling Theory (MCT) calculations to our system, the localization lengths normalized to radius should be around $7.46\cdot 10^{-2}$, our experiment, however, shows a localization length of $1.3 \cdot 10^{-3}$ to $4 \cdot 10^{-4}$ in backscattering and transmission geometry respectively\cite{sperl2005nearly,sperl2012single,Hunter2012Colloid}. It is remarkable that the MCT calculations agree with the Lindemann melting criterion or length, and still show a difference in MSD of 4 orders of magnitude from experimental data \cite{lindemann1910berechnung}. We suspect possible causes to be friction, roughnesses, or deviations from a sphere shape on the order of 1\%, as the mentioned MCT calculations are frictionless ideal spears. However, the core spectral characteristics of the MSD graph are evident in both theoretical and experimental contexts \cite{sperl2005nearly}.\\
\section{\label{sec:Conclusion}Conclusion}
We demonstrate that DWS can be used to produce reliable and reproducible MSDs from vibro-fluidized beds, in agreement with the literature for other types of fluidization, such as funnel fluidization and gas fluidization \cite{menon1997diffusing,menon1997particle}. All three regimes, ballistic, sub-diffusive, and diffusive, shown in previous granular DWS experiments can be observed in the present voice coil experiment. It is observed that the diffusive behavior changes to the sub-diffusive behavior of the system and undergoes a transition in response to agitation (granular temperature). In addition, we verify that the behavior of the system changes as it densifies, as a denser system slows down the dynamics. We expect links to jamming and a granular glass transition as the temperature is constant but the dynamics slow down. Compared to colloidal systems the mean free paths appear considerably smaller. Finally, we show that the dynamics observed in the backscattered and transmitted geometry are different, demonstrating the sensitivity of DWS to heterogeneities within the system, specifically between wall and bulk dynamics. Consequently, we conclude that DWS can provide a coherent picture of granular fluids across different agitation mechanisms, as it allows the measurement of particle dynamics independent of the agitation method. 

\section{Acknowledgment}
This work was supported by the DLR Space Administration with funds provided by the Federal Ministry for Economic Affairs and Climate Action (BMWK) based on a decision of the German Federal Parliament under grant number 50WM1945 (SoMaDy2). We thank for the discussions with Thomas Blochowicz, Matthias Schröter, Karsten Tell, Till Kranz, and Peidong Yu. 

\bibliography{VCbib}

\end{document}